\documentclass{article}
\usepackage[utf8]{inputenc}
\usepackage{amsmath,amssymb,amsfonts}
\usepackage[left=2cm,right=3cm]{geometry}
\usepackage[linesnumbered,ruled]{algorithm2e}
\usepackage{graphicx}
  \usepackage{authblk}
\usepackage{subfig}
\usepackage{float}
\usepackage{xcolor}

\begin{document}

\title{Danger of using fully homomorphic encryption:\\ A look at Microsoft SEAL}

\author{Zhiniang Peng}

\affil{Qihoo 360}
\affil{Email : jiushigujiu@gmail.com}

  \maketitle

%

\begin{abstract}
{Fully homomorphic encryption is a promising crypto primitive to encrypt your data while allowing others to compute on the encrypted data. But there are many well-known problems with fully homomorphic encryption such as CCA security and circuit privacy problem. Despite these problems, there are still many companies are currently using or preparing to use fully homomorphic encryption to build data security applications. It seems that the full homomorphic encryption is very close to practicality and these problems can be easily mitigated in implementation. Although the those problems are well known in theory, there is no public discussion of their actual impact on real application. Our research shows that there are many security pitfalls in fully homomorphic encryption from the perspective of practical application. The security problems of a fully homomorphic encryption in a real application is more severe than imagined. In this paper, we will take Microsoft SEAL as an examples to introduce the security pitfalls of fully homomorphic encryption from the perspective of implementation and practical application.}
\end{abstract}

\section{Introduction}

Fully homomorphic encryption is a promising crypto primitive to protect data while allowing others to compute on the data. But there are also many well-known problems with fully homomorphic encryption such as CCA security and circuit privacy problem. Despite these problems, there are still many companies are currently using or preparing to use fully homomorphic encryption to build data security applications. It seems that the full homomorphic encryption is very close to practicality and these problems can be easily mitigated in implementation. Although the those problems are well known in theory, there is no public discussion of their actual impact on real application. Our research shows that there are many security pitfalls in fully homomorphic encryption from the perspective of practical application. The security problems of a fully homomorphic encryption in a real application is more severe than imagined. In this paper, we will take Microsoft SEAL as an examples to introduce the security pitfalls of fully homomorphic encryption from the perspective of implementation and practical application.

Microsoft have published the source code of Microsoft Simple Encryption Math Library (Microsoft SEAL)\cite{SEAL}. SEAL aims to provide a high-performance and easy-to-use homomorphic encryption algorithm library. It has been adopted in several projects including the Intel Neural Network Compiler nGraph. Many companies are currently using or preparing to use SEAL to build data security applications based on fully homomorphic encryption. Therefore, we analyzed the security risks of SEAL and found out that there are many security pitfalls when building applications based by SEAL. And all of these security issues also apply to other fully Homomorphic Encryption libraries. Our research shows that the primitive (fully homomorphic encryption) itself is secure, protocols and applications can easily be completely insecure in actual use.

Outline of this article is as follows:
 \begin{enumerate}
     \item Introduction to homomorphic encryption and their applications.
     \item Introduction to BFV homomorphic encryption scheme in SEAL.
     \item Chosen-ciphertext attack on fully homomorphic encryption.
     \item Circuit privacy.
     \item Encoder information leakage.
     \item Other issues.
     \item Conclusion.
	\end{enumerate}

\subsection{Homomorphic Encryption}
With the popularity of Cloud computing, the demand for encrypted data searching and processing is increasing, making fully homomorphic encryption getting more and more attention. The idea of fully homomorphic encryption was first proposed by Rivest et al. in the 1970s. Compared with the general encryption algorithm, homomorphic encryption can let you do computation on ciphertexts.

An encryption function with homomorphism is an encryption function in which two plaintexts a and b satisfy $Dec (Enc(a)\bigoplus Enc(b))=a\bigotimes b$, where Enc stands for encryption, and Dec stands for decryption, $\bigoplus$, $\bigotimes$ corresponds to operations on the plaintext and ciphertext. When $\bigoplus $ represents addition on the plaintext, the encryption is said to be an additive homomorphism: when $\bigoplus$ represents multiplication on the plaintext, the encryption is said to be a multiplicative homomorphism.

A cryptosystem that supports arbitrary computation on ciphertexts is known as fully homomorphic encryption (FHE). That is, $Dec(f(Enc(m_1), Enc(m_2),..., Enc(m_k)))=f(m_1, m_2,...,m_k)$, or written as: $f(Enc(m_1) , Enc(m_2),...,Enc(m_k))=Enc(f(m_1,m_2,...,m_k))$, for arbitrary function $f$.

How to construct a fully homomorphic encryption scheme is an open challenge. Until 2009, Gentry proposed the first fully homomorphic cryptosystem based on the ideal lattice\cite{Gentry}, which made a breakthrough in this field. Then many cryptographers have done meaningful work in the research of the homomorphic encryption scheme.

\subsection{Application of Homomorphic Encryption}

With the popularity of Cloud computing, more and more user’s data are stored on Cloud servers. This raises concerns about privacy. If the user encrypts the data and then stores it in the cloud, then the privacy problem can be solved. However, using the traditional encryption scheme will result in Cloud not being able to process the data, or provide services to users. Homomorphic encryption can solve this problem. If the user encrypts the data using homomorphic encryption, the Cloud can perform meaningful computation on the encrypted data in order to provide services to users without revealing the user's privacy.

\subsection{Categories of Homomorphic Encryption}
In \cite{CAN}, authors presented some practical scenarios of Homomorphic Encryption. We can divide them into two categories:
     \begin{enumerate}
     \item The first category is that: the data is private and but function to be calculated is public.
     \item The second category is that: both the data and the function to be calculated are private.
	\end{enumerate}

There are two examples:

     \begin{enumerate}
     \item Disease prediction (private data , public function): In a Cloud storage system of medical records, patients use Homomorphic Encryption to encrypt their blood pressure, heart rate, blood sugar, age, weight, gender, historical cases and other data, and then upload them to the Cloud. The Cloud compute the weighted average, standard deviation, or other statistical functions (such as logistic regression) of these data to help users predict the probability of diseases in the future. The encrypted predicted value is then returned to the user, and the user decrypts it to obtain the corresponding result.

     \item Business forecasting (private data, private function): Company A has a large amount of enterprises’ private data (business operations, financial information, personnel, etc.), and Company B has a private business forecasting model. Under normal circumstances, without leaking data or the prediction model, A and B cannot predict the operation status of these companies. If A encrypts the data using homomorphic encryption and then sends it to B to evaluate the function to get the encrypted prediction result, then B returns the encrypted result to A. Now A and B can complete the prediction without revealing the data and the prediction model.
	\end{enumerate}

The first case can be considered outsourced compute. It is sufficient to require Homomorphic Encryption to be semantically secure. In the second case, it is can be considered a case of secure multiparty computation. To ensure the privacy of the functions, the homomorphic encryption must satisfy “circuit privacy” at the same time, otherwise the security of the scheme cannot be guaranteed.

\subsection{Practicality}
Microsoft's SEAL Homomorphic Encryption library implements two homomorphic cryptographic schemes, BFV\cite{BFV} scheme and CKKS scheme. BFV is adopted by in many applications. In CryptoNets\cite{Cryptonets}, Microsoft uses SEAL to perform neural network recognition (MIST handwritten digital picture recognition) on encrypted pictures. The throughout can reach about 60,000 times of encrypted picture recognition per hour for a single CPU while guaranteeing 99\% correct rate. More than 16 encrypted pictures are recognized per second in average. Intel also integrates SEAL in their neural network compiler nGraph, allowing artificial intelligence model to directly process encrypted data. It can use to analyze data without sacrificing data privacy.

In our tests, using SEAL to perform a logistic regression prediction on 10,000 pieces of data can be done in 5 minutes, which is about 300 times slower than using sk-learn directly on plaintext. It can be seen that in some scenarios with high privacy requirements, SEAL is already able to meet the demand in terms of performance.

However, in the use of SEAL, we found that it is very likely to meet security pitfalls while using SEAL to build application. Before talking about these pitfalls, we first briefly introduce how the BFV in SEAL works.

\section{BFV Homomorphic encryption in SEAL}
The BFV scheme is a homomorphic cryptographic scheme based on the Ring-LWE problem of ideal lattice. Details can be found in \cite{BFV}. The main parameters are as follows:

     \begin{enumerate}
     \item The polynomial $f(x) \in Z[x]$ is a d-degree irreducible polynomial, usually taking $f(x)=x^d+1$, where $d=2^n$.
     \item Polynomial ring $R=Z[x]/f(x)$.
     \item $q$, an integer greater than 1, which is the polynomial modulus. $Z_q$ represents a set of integers $(-\frac{q}{2}, \frac{q}{2})$, and $[a]_q$ is the congruence of $a$ mod $q$ in set $Z_q$.
     \item $t$ is an integer with $1 < t < q$ and is the plaintext modulus coefficient. $\delta =floor(\frac{q}{t})$, $Z_t$ represents a set of integers $(-\frac{t}{2}, \frac{t}{2})$ , and $[a]_t$ is the congruence of $a$ mod $q$ in set $Z_t$.
      \item $T$ is a positive integer. $l=floor(log(T,q))$
      \item $R_2$ is a random distribution: $R(randint(0,1)^d)$
      \item $R_e$ is the noise distribution: $R(D()^d)$, where $D()$ is a discrete Gaussian s	ample at over integer with standard deviation $\sigma$.
	\end{enumerate}

Since the original description of BFV scheme is too mathematical, here I wrote it down in pseudo-code:

\textbf{Generate the key:}
     \begin{enumerate}
     \item Randomly choose the private key $s\leftarrow R_2$
     \item Randomly choose noise $e\leftarrow R_e$
     \item Compute the public key  $pk[0]= [(-(a*s+e))]_q, pk[1]=a$. $[x]_q$ is used to calculate the congruence of $x$ mod $q$ in set $Z_q$
     \item Generate the relinearization key
	\end{enumerate}

\textbf{Encrypt(m):}
     \begin{enumerate}
     \item Randomly choose $u\leftarrow R$
     \item Randomly choose noise $e_1 \leftarrow R_e$
     \item Compute the ciphertext: $c[0]= [pk[0]*u+e1+\delta *m]_q$, and $c[1]= [pk[1]*u+e_2]_q$
	\end{enumerate}

\textbf{Decrypt(c):}
     \begin{enumerate}
     \item Compute: $tmp=[c[0]+c[1]*s]_q$
     \item Compute the message: $m=[(round(tmp*t/q)]_t$. $[x]_t$ is used to calculate the congruence of $x$ mod $q$ in set $Z_t$.
	\end{enumerate}

\textbf{Add($c_1$,$c_2$):} Compute ciphertext: $c[0]=[c_1[0]+c_2[0]]_q$, $c[1]=[c_1[1]+c_2[1]]_q$

Details of BFV scheme can be found in the original paper. We have implemented BFV scheme in sage, the source code can be found on our GitHub\footnote{https://github.com/edwardz246003/danger-of-using-homomorphic-encryption}. The implementation of BFV in SEAL uses the Chinese remainder theorem and the fast Fourier transform for the sake of performance. It is more complicated than our implementation. But the basic scheme is the same.

\section{Chosen-ciphertext attack: Recover the secret key with single decryption query}

\subsection{Fully homomorphic encryption does not meet IND-CCA security}
The BFV scheme has IND-CPA security, and its security is based on the D-RLWE problem. However, the BFV scheme does not have IND-CCA security. In fact, all current homomorphic encryption schemes that can be used in practice cannot meet the IND-CCA security. This problem had been discussed in many papers before \cite{CCAattack}. There are some research in the theoretical world such as \cite{CCAFHE} in PKC17, but the performance is completely impractical. Therefore, all solutions using the homomorphic encryption cannot guarantee security in the IND-CCA scenario.

\subsection{Scenarios that require Homomorphic Encryption often require IND-CCA security}

Although the the CCA problem of fully homomorphic encryption are well known in theory, there is no public discussion of their actual impact on real application. Our research shows that there are many security pitfalls in fully homomorphic encryption from the perspective of practical application. The security problems of a fully homomorphic encryption in a real application is more severe than imagined.
 
For example, in the scenario of outsourced computation, as long as the Cloud doesn’t have the opportunity to ask a decryption oracle, the application is running under the CPA model. However, there are always many rich data flows in real applications. It is difficult for data-owner to ensure that the decrypted data does not finally being leaked to the Cloud, thus breaking the CPA security model. In addition, scenarios that require homomorphic encryption usually involve multi-party’s cooperation and data exchange. It is very difficult to ensure that certain decrypted data is not leaked. Once the Cloud has access to additional information, the security of the homomorphic encryption scheme no longer exists.

Those who are counting on luck may think that although small range of leakage breaking the security model, it does not prove that the application is insecure. Well, in this section we will give an example that attacker can recover the whole secret key of the BFV scheme in just one decryption query.

While these attacks are very well known\cite{CCAattack}, much of the focus of the applied community has been in improving the performance in micro-benchmarks, not in creating secure systems from homomorphic encryption. Today, with better schemes and implementations available, this issue is urgent to be resolved.

\subsection{Chosen-ciphertext attack: recover the BFV scheme private key with a single decryption query}
In the BFV scheme, the key is $s$, and the process of decrypting ciphertext $c$ is:
     \begin{enumerate}
     \item Let: $tmp=[c[0]+c[1]*s]_q$
     \item Compute the message: $m=[round(tmp*t/q)_t$
	\end{enumerate}

If the attacker constructs ciphertext $c[0]=0$, $c[1]=\delta$. The decryption satisfies the following conditions:
     \begin{enumerate}
     \item $tmp=[0+\delta *s]_q$
     \item $m=[round(\delta *s*t/q)]_t=s$
	\end{enumerate}

\begin{figure}
\centering
  \includegraphics[scale=0.5]{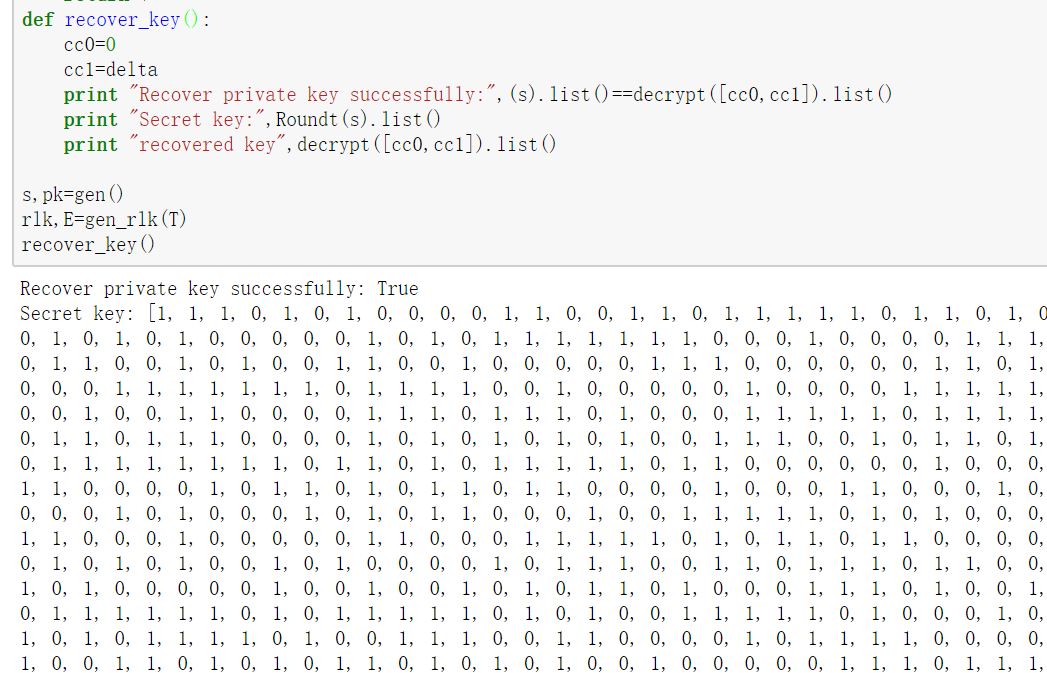}
    \caption{CCA attack on BFV}
    \label{NEO1}
\end{figure}

$\delta$ just happens to cancel with $t/q$ here. Finally, the plaintext $m=s$ is obtained. We implemented this attack on the recommended parameters in SEAL. Where $q=2^{54}, d=1024, t=256, T=100$. The attack code can be found at our github\footnote{https://github.com/edwardz246003/danger-of-using-homomorphic-encryption}. The attack results are show in figure~\ref{NEO1}.

It can be concluded that attacker only needs to have one decryption query to recover the secret key. One query attack of fully homomorphic encryption is well known in the crypto community. However, some still people think that preventing other people getting the decryption oracle may solve this problem. Actually, preventing other people getting the decryption oracle can not solve the problem fundamentally. Any information leakage of the decrypted result will result in leakage of the private key. In the next section, we will give another example that: 1-bit leakage of the decryption result will leak 1-bit of secret key. After a few times leakage, the security of fully homomorphic encryption will drop dramatically. 

Think a about it: In real application, every action you take may leak some bit of information and this will result in key leakage. And you can hardly make sure there is no information leakage between data owner and others. When designing a scheme based on homomorphic encryption, it must be ensured that no one except the data owner can get any information on the decryption result. But how can we make sure there is no information leakage in real application?

\textbf{Analysis: }
The problem is that the two ciphertext components are not tied to each other, and a malicious actor can change one of them. For a correctly constructed ciphertext decryption should remove all traces of the secret key from the decryption, but if c1 is changed, the change will interact with the secret key and leave a trace that may be located in the message part of the decryption, hence leaking potentially all of the key bits in a decryption oracle setting. We also note that the leakage might be extremely subtle, e.g., upon receiving an unintelligible result from a computation service provider, the customer cancels a monthly subscription and leaks (with some probability) a bit of their secret key. Even if only a part of the secret key is leaked, there are lattice attacks that leverage partial leakage and thus the security level may be catastrophically reduced.

\textbf{Countermeasures:}
Do not use homomorphic encryption in any scenario that decrypted result may leak to evaluators. This is easier said than done though, and it may be necessary for library developers to create larger components with an API that prevents insecure data flow. The HomomorphicEncryption.org standardization project may also want to consider standardizing simple protocols rather than only the lowest level encryption primitives. There may also be ways of detecting whether a ciphertext is potentially malicious; indeed, a covert security model might be good enough if encryption libraries refuse to decrypt ciphertext that are observed to be potentially malicious.

\section{Circuit Privacy}

\subsection{Match PSI and Private Contacts}
Private Set Intersection (PSI) is a cryptographic technique that allows two parties to calculate the intersection of two private input sets without revealing any other information. In a PSI scheme, there is one sender and one receiver. The sender has a private input set X, and the receiver has a private input set Y. The ideal PSI scheme is able to calculate $X\cap Y$ and send the result to the receiver, without revealing any other information except $X\cap Y$. A very interesting PSI application is: private contact discovery. For example, in an end-to-end encrypted communication app (Whatsapp for example), user has an contact list in his mobile phone, and he want to know which friend in his contact list is using WhatsApp. Of course, users don't want to completely reveal their contact list to the WhatsApp server. Then we can use PSI to do the private contact discovery.

\subsection{Constructing a fast PSI scheme from Homomorphic Encryption}
In CCS17, Chen\cite{CHEN2017} proposed  homomorphic encryption to construct a secure and efficient PSI protocol under the semi-honest security model to solve these problems. We call it FPSI.

The basic PSI protocol is show in figure~\ref{NEO2}:
\begin{figure}
\centering
  \includegraphics[scale=0.4]{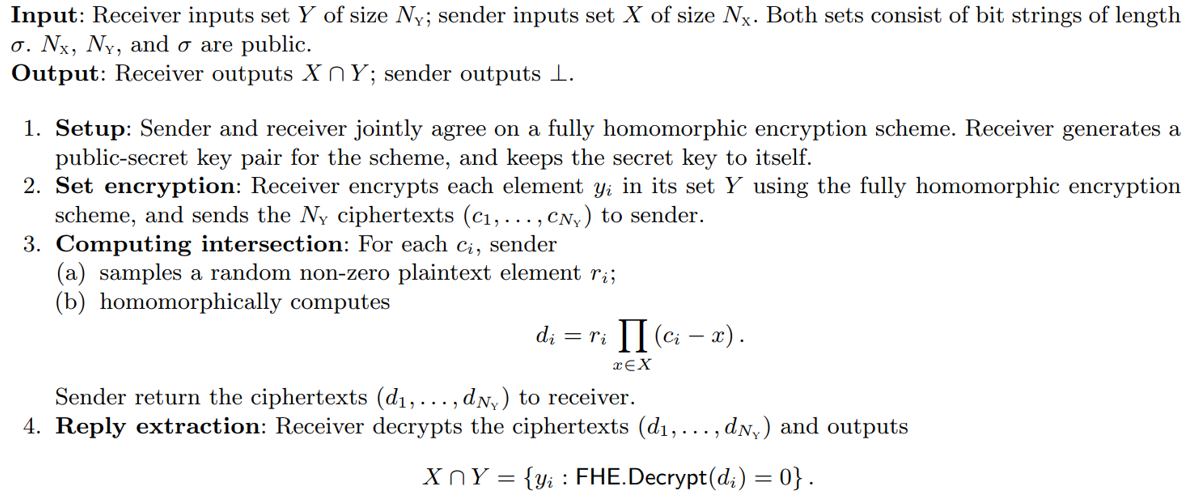}
    \caption{Basic PSI protocol.}
    \label{NEO2}
\end{figure}

A detailed security proof is given for the scheme in the paper. The proof shows that the FPSI is secured under the semi-honest model when the underlying full homomorphic encryption satisfies both IND-CPA and circuit privacy. SEAL is used to implement FPSI. The experiments show that the matching of 500 32-bit local strings and 16 million 32-bit server strings can be completed in about 100 seconds.

\subsection{The Security issues of FPSI}
From the perspective of the PSI scheme alone, proof has been given in the article. The PSI scheme is secure under the semi-honest model if the underlying homomorphic scheme satisfies both IND-CPA and circuit privacy conditions. But here we found two problems:
     \begin{enumerate}
     \item The scenario given by FPSI is private contact discovery. In this scenario, there is a possibility that the information (contact) decrypted by the user can be further leaked to the server (because user wants to add his friend in the app), then the security model for homomorphic encryption become CCA rather than CPA.
     \item The fully homomorphic encryption scheme in SEAL does not satisfy the circuit privacy. Without further processing, even the semi-honest users can recover the plaintext data of the server.
	\end{enumerate}

\subsection{Chosen-Ciphertext Attack against FPSI }
Now let’s consider the first issue. Assume that we use the FPSI scheme for private contact discovery on the end-to-end encrypted communication software. After user completes the protocol, he finds that phone numbers $x_i$ and $x_j$ in his address book are also registered in this app. Then user is likely to add $x_i$ and $x_j$ as friends through the app. At this time, $x_i$ and $x_j$ are actually being leaked to the server. It’s equivalent to that the server does a decryption query to the user. In this case, the server is able to carry out a chosen-ciphertext attack. Then it can recover the user's key and decrypt all the user's contact list.

But here it is not the same as the chosen-ciphertext attack model (one query attack) we mentioned above. Because rather than sending the decrypted result to the server, the user will check whether the decryption result is 0. If it is 0, it returns $x_i$ to the server; otherwise, it will not return. So, there is only 1 bit of information leakage per query. But here we will give an example that we can reveal 1 bit of user’s private key using this 1-bit information leakage.

     \begin{enumerate}
     \item Let: $M = \delta /4 + 20$, where 20 is a number large enough to cover up the noise.
     \item Let: $t_1=M*x^i, t_2=M$.
     \item Let ciphertext: $c_0=pk[0]+t_1,c_1=pk[1]+t_2$.
	\end{enumerate}

Then server submitted ciphertext $c=(c_0, c_1)$ to the user for decryption, and the decryption process is as follows:
\[c_0+c_1*s=pk[0]+t_1+(pk[1]+t_2)*s=-(as+e)+t_1+(a+t_2)*s=e+t_1+t_2*s\]

Since $t_1$ and $t_2$ are controlled by us, so the decryption result is all 0 except for the i-th bit, and the i-th bit is equal to the i-th bit of secret key s. Then, if the decryption result returned by the user is 0, the i-th information of the key is 0. If the returned result is not 0, the key i-th bit information is 1. (Here we assume that s has a value range of 0,1). We implemented this attack on the recommended parameters in SEAL, where $q=2^{54}$, $d=2048$, $t=256$, $T=100$. The attack code can be found at our Github\footnote{https://github.com/edwardz246003/danger-of-using-homomorphic-encryption}. The results are showed in figure~\ref{NEO3}:

\begin{figure}
\centering
  \includegraphics[scale=0.8]{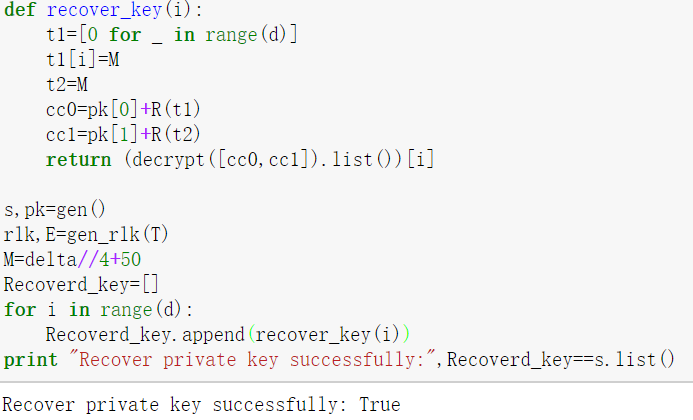}
    \caption{One bit information leakage result in one bit key leakage.}
    \label{NEO3}
\end{figure}

However, this attack needs the attacker actively generate malicious ciphertext. Therefore, it’s not a semi-honest attacker. But here we still would like to emphasize that when using homomorphic encryption, developers must thoroughly consider the application scenario; otherwise, it will easily lead to security problems. Chosen-ciphertext attacks may occur in many practical scenarios where fully homomorphic encryption schemes cannot prevent such attacks. The BFV scheme implemented in SEAL allows an attacker to recover a key by performing a decryption query. Even small information disclosure will result in the disclosure of the secret key.

\subsection{BFV circuit privacy analysis: data recovery against FPSI scheme}
As for the second issue, the homomorphic encryption in SEAL does not satisfy the circuit privacy if no additional security measures are taken, which will result in the private data being recover by the decrypting party. Although the authors of the FPSI program discussed possible security countermeasures in the Section 5.2 of the original paper, SEAL did not provide such an interface to the user. Below we will discuss the circuit privacy of the BFV scheme and demonstrate a practical attack against the FPSI scheme without noise flooding.

Take the simplest FPSI instance. Suppose Alice and Bob each choose a number $m_a$ and $m_b$, respectively. Alice wants to know if Bob’s number is equal to hers, so does Bob, without leaking $m_a$ and $m_b$ to each other. The workflow is as follows:

     \begin{enumerate}
     \item Alice generates a public-private key pair for the BFV scheme, then she sends the public key to Bob.
     \item Alice encrypts $m_a$ using the BFV scheme to get $c_a$ and sends it to Bob.
     \item Bob choose a random non-zero plaintext $r$ and then performs a ciphertext evaluation: $c_{ab}=r*(m_b-c_a)$ and sends the result to Alice.
     \item After Alice decrypts $c_{ab}$, she check whether the $c_{ab}$ is 0. If it is 0, then $m_a$ is equal to $m_b$. Otherwise, $m_a$ is not equal to $m_b$.
	\end{enumerate}

We chose the parameter $q=2^{54}$ $d=2048$, $t=83$ to run this protocol. The result is show in figure~\ref{NEO4}:
\begin{figure}
\centering
  \includegraphics[scale=0.8]{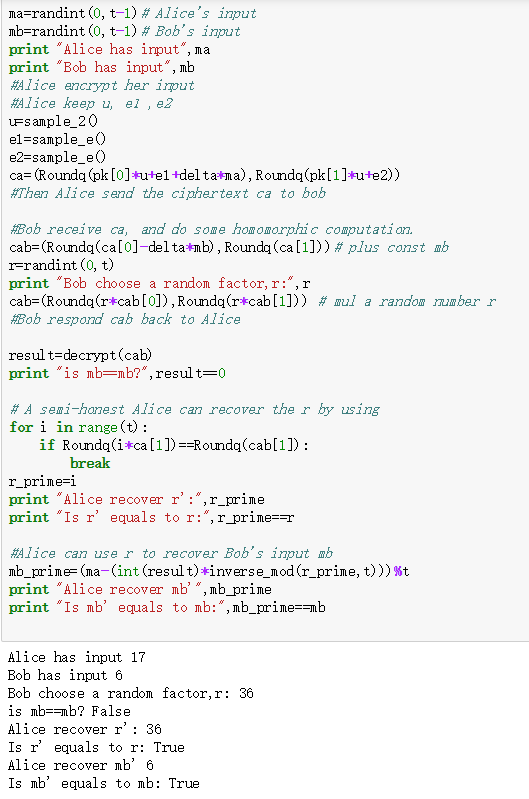}
    \caption{Circuit privacy of SEAL.}
    \label{NEO4}
\end{figure}

From the figure~\ref{NEO4}, we can see that Alice and Bob can use the FPSI scheme to finish the equality check. But Alice is able to recover Bob's plaintext message $m_b$. The essential reason is that in the SEAL, Bob performs mul\_const and add\_const operations without introducing any random factor, so Alice can directly recover $r$ and $m_b$.

Bob introduce randomness by encrypting $r$ and $m_b$ before performing the evaluation, Alice can still use noise $e_1$ and $e_2$ to compute the of amount of noise that Alice introduces after the evaluation, so as to infer the $r$ and $m_b$. Therefore, directly using SEAL to implement FPSI does not guarantee the security, additional countermeasures should be taken.

\textbf{Countermeasures:}
Circuit privacy issues are addressed in Section 9.4 of the SEAL Handbook\footnote{https://www.microsoft.com/en-us/research/wp-content/uploads/2017/06/sealmanual\_v2.2.pdf} and CCS17 FPSI. It is suggested that the best practice is "noise flooding", that is, adding an encrypted 0 to the final result. In the same time, ensure that the encrypted 0 contains enough noise to cover up the noise. However, there is no "noise flooding" interface provided in the SEAL, and designing such noise requires professional knowledge on lattice-based crypto, which is difficult for common users. We would like to note that noise flooding is a reasonable technique to use and does not cause significant overhead when used with larger encryption parameters. Libraries should provide a noise flooding and ciphertext re-randomization interface. By the way, how can the evaulator know how much noise he need to add? Is it also a information we should protect in some scenarios?

An improved PSI protocol\cite{CHEN2018} was published by Chen, Laine, and Rindal in CCS ’18; it has a malicious security property that protects against maliciously constructed ciphertexts and does not need circuit privacy. However, these stronger properties come from leveraging other cryptographic primitives in conjunction with homomorphic encryption, and building such protocols is far out of reach from normal software engineers.

\section{Encoder Information Leakage}
The plaintext space of homomorphic encryption is on a plaintext ring. However, in actual applications, the data we need to process may be floating point numbers, integers, and so on. So SEAL provides an encoding method that encodes integer or floating point numbers into the plaintext ring. However, it is not a one-to-one encoding, and it will reveal the data.

Here we consider a classic cryptography problem "The Millionaires’ Problem". Suppose there are three millionaires A, B and C. A has $m_1$ million dollar, B has $m_2$ million dollar, and C has $m_3$ million dollar, of which $m_1=1$, $m_2=3$, and $m_3=4$. They want to know the total amount of their assets, but they don't want the others to know how much each one have.

Here is a multiparty computation from homomorphic encryption to do that. This scheme is different from the standard scheme of MPC constructed by full homomorphism \(SPDZ\). It is mainly to explain the encoder information leakage issue in SEAL in a more simple way. SPDZ based construction is also vulnerable to the encoding problem described in this article.:

     \begin{enumerate}
     \item C generates a pair of public and private key of BFV, and sends the public key to A and B.
     \item A encodes $m_1=1$: $m_1'=IntegerEncoder(m_1)=1$. Then A encrypts $m_1'$ to get $c_1$=encrypt($m_1'$). A sends $c_1$ to B.
     \item B encodes $m_2=3$: $m_2'=IntegerEncoder(m_2)=x+1$. Then B encrypts $m_2'$ to get $c_2$ = encrypt($m_2'$). Then, compute $c_{12}$=add($c_1$, $c_2$) and send $c_{12}$ to C.
     \item C decrypts $c_{12}$ as decrypt($c_{12}$)=$x+2$. Decode $x+1$ result to $m_1+m_2=IntegerDecoder(x+2)=4$. Then C computes $sum(m_1+m_2+m_3)=8$.
      \item C sends the result 8 to A and B. Everyone knows the total amount of money, but no one knows how much money each person has.
	\end{enumerate}

\begin{figure}
\centering
  \includegraphics[scale=1]{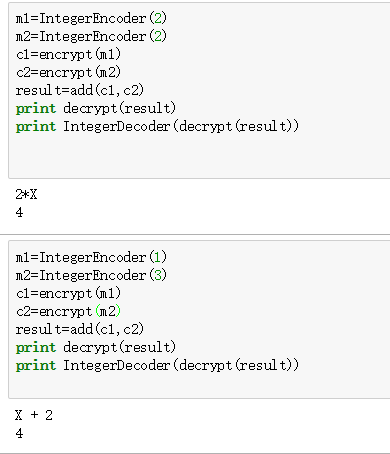}
    \caption{Encoder information leakage in SEAL.}
    \label{NEO5}
\end{figure}

But in fact, SEAL’s encoding will leak information of the operand. So, C can compute how much A and B actually have. As you can see in figure~\ref{NEO5}, different operand result in different encoding in the end, although the decoded integer is the same.

Ideally, C should only know that $m_1+m_2=4$, without knowing if it is $(m_1=1, m_2=3)$ or $(m_1=1, m_2=3)$. Although either $(m_1=1, m_2=1)$ or $(m_1=2, m_2=3)$, the final solution of $m_1+m_2$ is 4. But in fact, the result of decrypt($c_{12}$) is different.
     \begin{enumerate}
     \item If $(m_1=2, m_2=3)$. Then decrypt($c_{12}$)=$x+2$.
     \item If $(m_1=1, m_2=1)$. Then decrypt($c_{12}$)=$2x$.
	\end{enumerate}

\textbf{Analysis:}
The crypto part of fully homomorphic encryption library may have provable security, but the encoder part doesn't. Integer encoding in SEAL is many-to-one, which will leak information of operand. Technique like “noise flooding” cannot solve this problem in SEAL. Similar coding problems may exist in other fully homomorphic encryption libraries.

\textbf{Countermeasures:}
We have not found this issue with in BatchEncoder of SEAL; users can use it to prevent this problem, whereas the IntegerEncoder should be considered primarily as a demonstrative tool, rather than something that should be used in a real application.

\section{Other Issues}
Although the homomorphic encryption algorithm is an encryption algorithm, it indeed does not provide the security features as the commonly known encryption algorithms. For example, full homomorphic encryption is not an Authenticated Encryption. If the data is encrypted and stored directly using fully homomorphic encryption, it cannot guarantee the integrity of the data. An attacker can use the homomorphic nature of full homomorphic encryption to modify the ciphertext. Therefore, full homomorphic encryption cannot be directly used for encrypted storage and encrypted data transmission. Microsoft is currently leading the development of security standards for fully homomorphic encryption, but the latest version of the standard documentation still lacks a discussion of the security risks of using fully homomorphic encryption.

\section{Conclusion}

In this paper we described some of the security pitfalls of using the Microsoft SEAL. We found that SEAL is highly efficient and easy to use. But it is not as perfect as we originally thought. The scenario should be limited to pure outsourced computing. That is "all data comes from one party and the other does not have any privacy to protect". It is easy to generate security problems when it is used to solve other needs. All of these security issues also apply to other fully homomorphic encryption libraries. This confirms the seriousness of the security problems that fully homomorphic encryption will have in actual use.  While the primitive (fully homomorphic encryption) itself is secure, protocols and applications can easily be completely insecure. This is definitely something the entire community needs to work on going forward.

\section{Acknowledgement}
Author would like to thank Hong Cheng (Senior Expert of Alibaba Gemini Lab) and Kim Laine (Researcher of Microsoft Research) for their help with this paper.



\end{document}